\documentclass{llncs}

\newif\ifdraft
\draftfalse

\usepackage[T1]{fontenc}
\usepackage[utf8]{inputenc}
\usepackage{lmodern}
\usepackage[scaled=.8]{beramono}
\usepackage{textcomp}
\ifdraft
\usepackage[show]{ed/ed}
\else
\usepackage[hide]{ed/ed}
\fi

\usepackage[hyperref=auto,firstinits=true]{biblatex}
\usepackage[babel]{csquotes}
\MakeAutoQuote{“}{”}
\MakeAutoQuote*{‘}{’}
\usepackage{paralist}
\usepackage{graphicx}
\usepackage{lstsemantic}
\usepackage{url}

\begin{document}

\title{The Distributed Ontology Language (DOL):\\ Ontology Integration and Interoperability\\ Applied to Mathematical Formalization\thanks{The development of DOL is supported by the German Research Foundation (DFG), Project I1-[OntoSpace] of the SFB/TR 8 “Spatial Cognition”; the first author is additionally supported by EPSRC grant EP/J007498/1.  The authors would like to thank the OntoIOp working group within ISO/TC 37/SC 3 for their feedback.  The final publication is available at \texttt{http://www.springerlink.com}.}}
\author{%
Christoph Lange\inst{1,2}
\and Oliver Kutz\inst{1} 
\and Till Mossakowski\inst{1,3}
\and Michael Grüninger\inst{4}
}
\institute{%
Research Center on Spatial Cognition, University of Bremen%
\and Computer Science, University of Birmingham%
\and DFKI GmbH Bremen%
\and Department of Mechanical and Industrial Engineering, University of Toronto\\
\url{http://ontolog.cim3.net/cgi-bin/wiki.pl?OntoIOp}
}

\maketitle

\begin{abstract}
  The Distributed Ontology Language (DOL) is currently being standardized within the OntoIOp (Ontology Integration and Interoperability) activity of ISO/TC 37/SC 3.  It aims at providing a unified framework for \begin{inparaenum}[(1)]
  \item ontologies formalized in heterogeneous logics,
  \item modular ontologies,
  \item links between ontologies, and
  \item annotation of ontologies.
  \end{inparaenum}
  This paper focuses on an application of DOL's meta-theoretical features in mathematical formalization: validating relationships between ontological formalizations of mathematical concepts in COLORE (Common Logic Repository), which provide the foundation for formalizing real-world notions such as spatial and temporal relations.
\end{abstract}

\section{Distributed Ontologies for Interoperability}
\label{sec:intr--distr}

An ontology is a formal description (in a logical language) of the concepts and relationships that are of interest to an agent (user or service) or a community of agents.  Today, ontologies are applied in virtually all information-rich endeavors, for example eBusiness, eHealth, eLearning, and ambient assisted living.  Ontologies facilitate semantic integration of data and services by providing a common formal model, onto which data from different sources, as well as descriptions of different services, can be mapped.

Complex applications, which involve multiple ontologies with overlapping concept spaces, also require data mapping on a higher level of abstraction, viz.\ between different ontologies, where it is called ontology alignment.  While ontology alignment is most commonly studied for ontologies in the same logic, the different ontologies driving complex applications may also be formalized in \emph{different} logics.  Popular choices include propositional logic (e.g.\ in industrial requirements engineering), description logic (e.g.\ in biomedical applications and semantic web services), and first-order logic (required for formalizing mereology and notions of space and time, but exhibiting undecidable reasoning tasks).  Our approach faces this diversity not by proposing yet another ontology language – based on a logic that would subsume all the others – but instead we \emph{accept the diverse reality} and formulate means (on a sound and formal semantic basis) to \emph{compare and integrate ontologies formalized in different logics}.  We aim at addressing the challenge of automatically checking the coherence (e.g.\ consistency, conservativity, intended consequences) of ontologies and ontology-based services.

\section{The Distributed Ontology Language (DOL) – Overview}
\label{sec:distr-ontol-lang}

An ontology in the Distributed Ontology Language (DOL) consists of modules formalized in basic ontology languages, such as OWL (based on description logic) or Common Logic (based on first-order logic with some second-order features).  These modules are serialized in the existing syntaxes of these languages as to facilitate reuse of existing ontologies.  DOL adds a meta-level on top, which allows for expressing heterogeneous ontologies and links between ontologies.\footnote{The languages that we call “basic” ontology languages here are usually limited to one logic and do not provide meta-theoretical constructs.}  Such links include (heterogeneous) imports and alignments, conservative extensions (important for the study of ontology modules), and theory interpretations (important for reusing proofs).  Thus, DOL gives ontology interoperability a formal grounding and makes heterogeneous ontologies and services based on them amenable to automated verification.

DOL is currently being standardized within the OntoIOp (Ontology Integration and Interoperability) activity of ISO/TC 37/SC 3\footnote{TC = technical committee, SC = subcommittee}.  The international working group comprises around 50 experts (around 15 active contributors so far), representing a large number of communities in ontological research and application, such as different \begin{inparaenum}[(1)]
\item ontology languages and logics (e.g.\ the Common Logic and OWL),
\item conceptual and theoretical foundations (e.g.\ model theory),
\item technical foundations (e.g.\ ontology engineering methodologies and linked open data), and
\item application areas (e.g.\ manufacturing).
\end{inparaenum}  For details and earlier publications, see the OntoIOp project page%~\cite{OntoIOp}
.

The OntoIOp/DOL standard is currently in the working draft stage and will be submitted as a committee draft (the first formal ISO standardization stage) in August 2012.\footnote{The standard draft itself is not %currently
publicly available, but negotiations are under way to make the final standard document public, as has been done with the related Common Logic standard~\cite{CommonLogic:biblatex}.}  The final international standard ISO 17347 is scheduled for 2015.  The standard specifies syntax, semantics, and conformance criteria:
\begin{description}
\item[Syntax:] abstract syntax of distributed ontologies and their parts; three concrete syntaxes: a text-oriented one for humans, XML and RDF for exchange among tools and services, where RDF particularly addresses exchange on the Web.
\item[Semantics:] \begin{inparaenum}[(1)]
  \item a \emph{direct set-theoretical semantics} for the core of the language, extended by an \emph{institutional and category-theoretic semantics} for advanced features such as ontology combinations (technically co-limits), where basic ontologies keep their original semantics;
  \item a \emph{translational semantics}, employing the semantics of the expressive Common Logic ontology language for all basic ontologies, taking advantage of the fact that for all basic ontology languages known so far translations to Common Logic have been specified or are known to exist\footnote{Even for higher-order logics this works, in principle, by using combinators.};
  \item finally, there is the option of providing a \emph{collapsed semantics}, where the semantics of the meta-theoretical language level provided by DOL (logically heterogeneous ontologies and links between them) is not just specified on paper in semiformal mathematical textbook style, but once more formalized in Common Logic, thus in principle allowing for machine verification of meta properties\end{inparaenum}.  For details about the semantics, see~\cite{MLK:3SemanticsForDistributedOntologies12}.
\item[Conformance criteria] provide for DOL's extensibility to other basic ontology languages than those considered so far, including possible future languages.  \begin{inparaenum}[(1)]
  \item A \emph{basic ontology language} conforms with DOL if its underlying logic has a set-theoretic or, for the extended DOL features, an institutional semantics.  Similar criteria apply to translations between languages.
  \item A concrete syntax (\emph{serialization}) of a basic ontology language conforms if it supports IRIs (Unicode-aware Web-scalable identifers) for symbols and satisfies some further well-formedness criteria.
  \item A \emph{document} conforms if it is well-formed w.r.t.\ one of the DOL concrete syntaxes, which particularly requires explicitly mentioning all logics and translations employed.
  \item An \emph{application} essentially conforms if it is capable of processing conforming documents, and providing logical information that is implied by the formal semantics.
  \end{inparaenum}
\end{description}

\begin{figure}
  %\centering
  %% trim: left bottom right top
  \includegraphics[trim=0cm 0.5cm 10cm 2.1cm,clip,width=\textwidth]{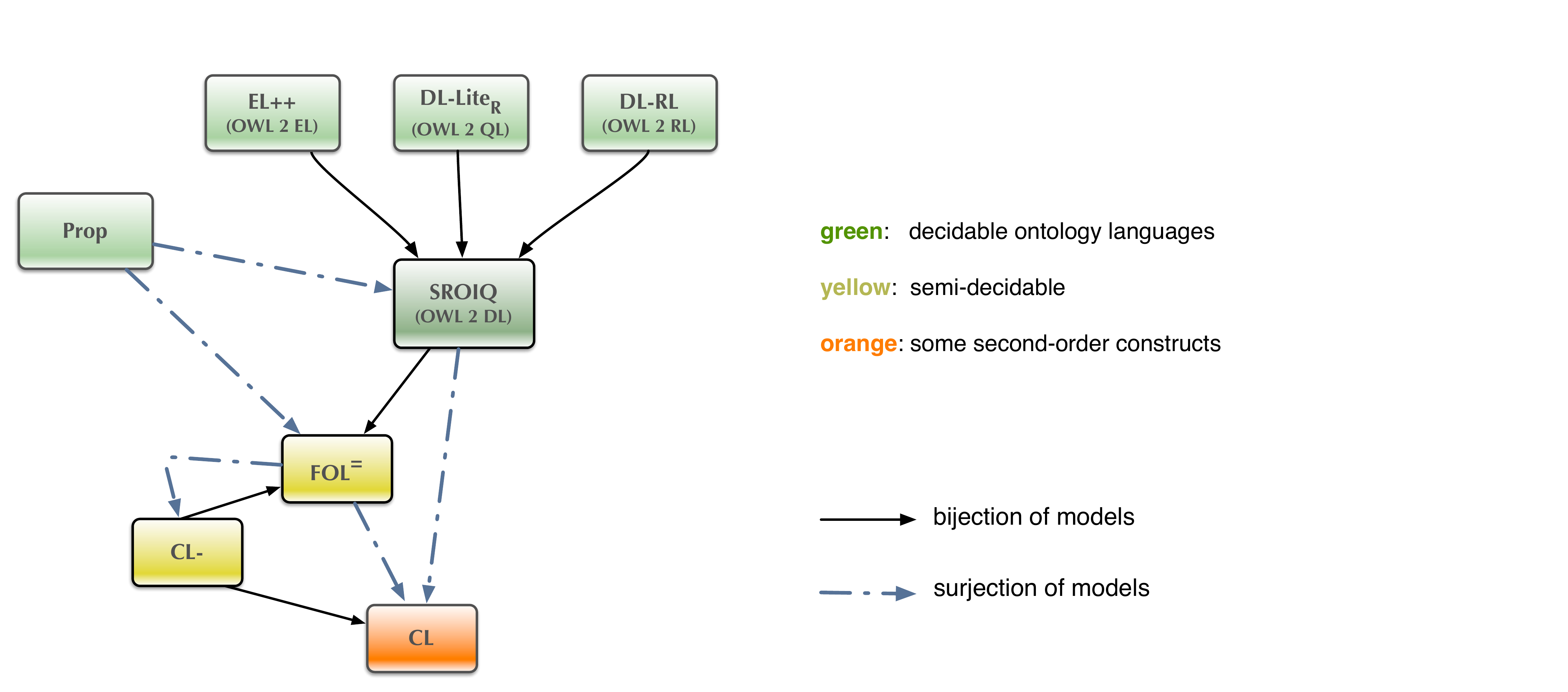}
  \caption{A core logic translation graph for basic ontology languages}
  \label{fig:coregraph}
\end{figure}

Figure~\ref{fig:coregraph} shows some logics and translations relevant for ontologies\footnote{The logics and translations have mostly been specified as part
of the Logic Atlas~\cite{CodHorKoh:palai11}.}: $\mathcal{SROIQ}$, the logic of OWL, and its sublogics corresponding to sublanguages (“profiles”) of OWL – $\mathcal{EL}\mathrm{++}$, DL-Lite$_R$, and RL – which aim at lowering the complexity of reasoning in certain relevant application domains; propositional logic; first-order logic with equality, Common Logic without second-order features, and full Common Logic.  We have defined the translations between them in earlier publications~\cite{MLK:3SemanticsForDistributedOntologies12,MoKu:OntoGraph11%:short
}; elaborating these definitions into annexes to the standard%% DONE
%\ednote{CL: Maybe some won't become annexes, as discussed in the 2012-02-23 meeting, but we need not go into such details here.}
, which establish the conformance of these logics and translations with DOL, remains to be done.

\section{Applications to Mathematical Formalization}
\label{sec:applications}

The first application of DOL can be seen in COLORE, the Common Logic Repository, an open repository of more than 500 Common Logic ontologies.  The objective of COLORE is to provide an “adequate set of generic ontologies that can be used to specify the semantics of primitive concepts”, as, for example, “any product ontology must refer to relationships from geometry and topology, and different manufacturing standards may require different ontologies for time”\footnote{\url{http://colore.googlecode.com}}.  One of the primary applications of COLORE is to support the verification of ontologies for commonsense domains such as time, space, shape, and processes.  Verification consists in proving that the ontology is equivalent to a set of core ontologies for mathematical domains such as orderings, incidence structures, graphs, and algebraic structures.   COLORE comprises core ontologies that formalize algebraic stuctures (such as groups, fields, and vector spaces), orderings (such as partial orderings, lattices, betweenness), graphs, and incidence structures in Common Logic, and, based on these, representation theorems for generic ontologies for the above-mentioned commonsense domains.

Meta-theoretical relationships between these ontologies are of particular interest, including maps (signature morphisms), definitional extension, conservative extension, inconsistency between modules, imports, relative interpretation, faithful interpretation, and definable equivalence.  DOL allows for formalizing them (as compared to the earlier approach of just writing them down as metadata), and we have started to automatically verify them using Hets (Heterogeneous Tool Set~\cite{MosMaeLue:thts07%:short
}).  The listing below shows an example\footnote{An excerpt from \url{https://colore.googlecode.com/svn/trunk/ontologies/complex/owltime/owltime_interval/mappings/owltime_le.dol}; the individual ontologies are actually stored in separate files, but here we demonstrate DOL's ability to maintain different ontologies within one file.} for interpreting linear orders (\url{linear_ordering}) as orders between time intervals that begin and end with an instant (\url{owltime_le}).  A third ontology (\url{mappings/owltime2orderings}) takes care of mapping the different predicate names used by the source and the target ontology, respectively.  We state that the source ontology can be interpreted in terms of the union of the target ontology and the mapping ontology in a model-theoretically conservative way, and that \url{mappings/owltime2orderings} extends
\url{owltime_le} with definitions.  
%The latter annotations \textbf{CL@TM: explain the benefits of having them vs.\ not having them} ….

\begin{lstlisting}[basicstyle=\ttfamily\scriptsize,language=dolText,morekeywords={props,ObjectProperty,Class,DisjointUnionOf,SubClassOf,Characteristics,Transitive,Asymmetric,SubPropertyOf,DisjointClasses,EquivalentTo,inverse,only,cl-imports,forall,iff,if,or,exists},escapechar=@,mathescape]
%prefix(                              %% prefixes for abbreviating long IRIs: this distributed ontology
 :    <http://code.google.com/p/colore/.../owltime/owltime_interval/mappings/owltime_le.dol#>
 log: <http://purl.net/dol/logics/>                                   %% DOL-conforming logics@ (Fig.~\ref{fig:coregraph})@
 ser: <http://purl.net/dol/serializations/>                   %% serializations, i.e. concrete syntaxes
 int: <http://code.google.com/p/colore/.../owltime/owltime_interval/>   %% namespaces of the ontologies
 ord: <http://code.google.com/p/colore/.../orderings/> )%               %% in this distributed ontology

%% The following ontologies are in the logic Common Logic, and written in the Lisp-style CLIF syntax
logic log:CommonLogic syntax ser:CommonLogic/CLIF

ontology ord:linear_ordering =       %% Here we use Common Logic's ontology import facility, but ...
  (cl-imports ord:partial_ordering) (forall (x y) (or (leq x y) (leq y x) (= x y)))

%% DOL also has a general import facility: We create the ontology of linearly ordered time intervals
ontology int:owltime_le = int:owltime_linear then int:owltime_e %% that begin and end with an instant 
%% ... by extending linearly ordered time intervals with intervals that begin and end with an instant

ontology int:mappings/owltime2orderings = (forall (x y) (iff (leq x y) (or (before x y) (= x y))))
  (forall (x y) (iff (lt x y) (before x y)))          %% map time intervals to general linear orderings

interpretation i %mcons :                      %% interpreting linear orderings as time interval orders
  ord:linear_ordering to {int:owltime_le and %def int:mappings/owltime2orderings}
\end{lstlisting}

DOL can also be used to specify the relationships between ontologies axiomatized in different logics. There are several cases in which 
there exist ontologies for the same domain, some of which are axiomatized
in description logic and others in first-order logic.
The best example of this is OWL-Time, which was originally proposed with
an OWL axiomatization, and later extended with a first-order axiomatization
\cite{hobbspan04}.  (COLORE includes a modularized version of OWL-Time~\cite{Gruninger:VerificationOWLTime11}; the listing shows some of the modules.)  Using DOL, one can specify that the first-order axiomatization
is a nonconservative extension of the OWL axiomatization, but that there 
exists a subtheory of the first-order axiomatization that is definably
equivalent to the OWL axiomatization.

% kwarc.bib contains all that I (Christoph) am interested in.  Let's add new references to cicm2012.bib.  Additionally please copy any relevant *.bib files here.
\printbibliography
\end{document}